\documentstyle[prl,aps,epsfig,multicol]{revtex}
\draft
\begin{document}
\title{Nonequilibrium Campbell length: probing the vortex pinning potential}
\author{R. Prozorov and R. W. Giannetta}
\address{Loomis Laboratory of Physics, University of Illinois at Urbana-
Champaign, 1110 West Green Street, Urbana, Illinois 61801.}
\author{T. Tamegai}
\address{Department of Applied Physics, The University of Tokyo, Hongo,
Bunkyo-ku, Tokyo, 113-8656, Japan.}
\author{P. Guptasarma and D. G. Hinks}
\address{Chemistry and Materials Science Division, Argonne National
Laboratory, Argonne, Illinois 60439.}
\date{Submitted April 4, 2000}
\maketitle

\begin{abstract}
The $AC$ magnetic penetration depth $\lambda (T,H,j)$ was measured in
presence of a macroscopic $DC$ (Bean) supercurrent, $j$. In single
crystal BSCCO below approximately 28 K, $\lambda (T,H,j)$ exhibits
thermal hysteresis. The irreversibility arises from a shift of the vortex
position within its pinning well as $j$ changes. It is demonstrated that
below a new irreversibility temperature, the nonequilibrium Campbell
length depends upon the ratio $j/j_c$. $\lambda (T,H,j)$ {\it increases}
with $j/j_c$ as expected for a non-parabolic potential well whose
curvature {\it decreases} with the displacement. Qualitatively similar
results are observed in other high-$T_{c}$ and conventional
superconductors.
\end{abstract}


\begin{multicols}{2}
\narrowtext

The $AC$ penetration depth $\lambda$ is an extremely sensitive probe of
the vortex state. Models for the small-amplitude $AC$ response usually
assume a uniform distribution of vortices and a parabolic effective
potential well \cite{campbell,clem,brandt,koshelev,beek,blatter}.
However, experiments often probe a nonequilibrium flux profile. In this
case, $\lambda$ may depend not only upon $H$ and $T$ but also upon the
Bean supercurrent $j$ arising from a gradient of vortex density. In this
paper we show that a finite $j$ produces a hysteretic $AC$ response
whenever the pinning potential is non-parabolic. In contrast to most
experiments, the irreversibility discussed here occurs with temperature
while the applied field is held constant. Thermal hysteresis is observed
in several materials, but we focus on BSCCO. In all our experiments
$H_{ac}$ ($\sim 5$ mOe) $\ll H_{dc} \ll H_{c2}$ where the excitation
field does not modify the average vortex distribution and we can ignore
vortex core contribution to $\lambda$.

Vortices transmit the perturbation caused by a small $AC$ field as either
compressional or tilt waves, depending upon the geometry of the
experiment \cite{brandt}. If the amplitude of the $AC$ field is small,
the elastic restoring force is: $F=-\alpha u$, where $u$ is the vortex
displacement caused by a small $AC$ current and $\alpha$ is the Labusch
parameter. The latter incorporates, self - consistently, both the pinning
and the elastic forces. The $AC$ penetration depth is then given by
$\lambda^2 =\lambda_{L}^2+\lambda_{C}^2$ where $\lambda_{C}$ is the
Campbell penetration depth \cite{campbell,clem,brandt,koshelev}. The
Campbell length is $ \lambda _{C}=\sqrt{C_{xx}/\alpha}$, where $C_{xx}$
is the appropriate elastic modulus ($C_{11}$ for compression or $C_{44}$
for tilt, $C_{44} \approx C_{11} \approx B^2/4\pi$)
\cite{brandt,koshelev}. Unlike earlier versions based on the local
pinning force \cite{campbell}, these formulae are also good
approximations for a non-local elastic response with dispersive $\alpha
(k)$ \cite{brandt}. It is also possible to incorporate flux creep and
flux flow effects \cite{clem,brandt,koshelev,beek,blatter}. These
effects, however, do not result in thermal irreversibility of $\lambda
(T,H,j)$. In this paper, we explore the simplest generalization of the
elastic model. In the presence of a macroscopic $DC$ Bean current, $j$,
the Lorentz force biases the mean vortex position away from the pinning
potential minimum. The Campbell depth is then determined by the small -
amplitude response near this new bias point. If the dependence of the
pinning potential upon vortex displacement $u$ is non-parabolic, then
changes in either $j$ or the critical current $j_c$ will result in
changes of the Campbell length. If $j$ relaxes via flux creep or $j_c$
changes with temperature, the Campbell length will then show a small, but
observable thermal hysteresis.

\begin{figure}
 \centerline{\psfig{figure=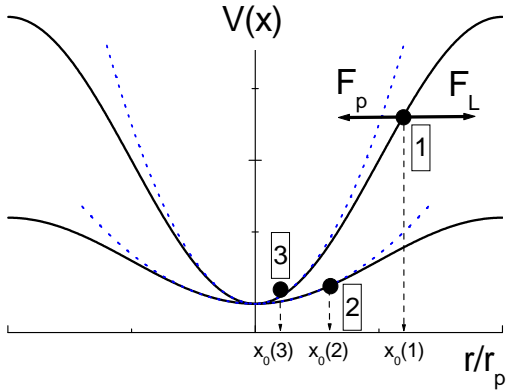,width=7cm,clip}}
 \caption{Schematic illustration of the effective pinning potential.
 Solid and dotted lines depict non-parabolic and parabolic potentials,
respectively. Filled circles represent vortex position in different
situations described in the text.}
 \label{fig1}
\end{figure}

Let $V(x)$ describe the shape of the pinning potential with $x=r/r_p$,
where $r$ is the vortex displacement and $r_p$ is the effective pinning
radius, $r_p \sim \xi$ for a quenched disorder pinning at low
temperatures ( see Fig. \ref{fig1}). The Lorentz force due to $j$ causes
a dimensionless vortex displacement $x_0$ determined from the condition:
$dV/dx=Bj/r_p c=\alpha_{0}(j/j_c)$, where $\alpha_0 \equiv Bj_c/c r_p$.
Upon application of a small $AC$ field, the restoring force acting on a
vortex in the vicinity of $x_0$ is $F(u) \simeq - \alpha u + O(u^2)$
where $u=x-x_0$. It is assumed that $j$ does not change during an $AC$
field cycle. The restoring elastic constant is then
$\alpha=d^2V/dx^2|_{x=x_0}$. For a parabolic potential,
$\alpha=\alpha_0=const$ and does not depend on $j$. For a more general
form of the potential we take into account a cubic term in $V(x)$ in the
vicinity of $x_0$. A Labusch parameter obtained from the Taylor expansion
of the restoring force $F=-dV/dx$ around $x_0$ can be written as
$\alpha=\alpha_0(1-\beta x_0)$, where $\beta$ is a dimensionless
constant. Positive $\beta$ describes pinning potential $V(x)$ saturating
at large $x_0$, while negative $\beta$ corresponds to a potential whose
curvature increases with the increase of $x_0$. The saturation of a
volume pinning force, corresponding to $\beta
> 0$, was discussed by Brandt on the basis of the results of numerical
simulations \cite{brandt}. Saturation also holds for weak collective
pinning. A saturating potential provides a smooth crossover to the flux
flow regime, where the penetration depth diverges (for zero viscosity).
Negative $\beta$ could be realized for pinning by columnar defects.
Assuming $\left| \beta x_0 \right| \ll 1$ (small deviation of $V(x)$ from
parabolic form) we obtain for the vortex displacement due to current $j$:
$x_0 \approx j/j_c$, which gives $\alpha=\alpha_0 (1 - \beta j/j_c)$. The
linear $AC$ response in the presence of a Bean current $j$ is then
described by:
\begin{equation}
\lambda^2 \left( j \right) \simeq {\lambda _L^2 + \frac{\lambda
_{C}^2 \left( 0 \right)}{1 - \beta j/j_c }} \label{lacj}
\end{equation}

If pinning is weak and magnetic relaxation is fast, as is typical for
high-$T_c$ superconductors \cite{blatter}, the term $\beta j/j_c$ is
small. Equation (\ref{lacj}) then reduces to:
\begin{equation}
\Delta \lambda \equiv \lambda \left( j \right) - \lambda \left( 0
\right) = \beta\frac{\lambda _C^2 \left( 0 \right)}{2\lambda
\left( 0 \right)}\frac{j}{j_c} \label{deltalacj}
\end{equation}

Equations (\ref{lacj}) and (\ref{deltalacj}) predict that a non-parabolic
potential results in an explicit dependence of $\lambda$ on $j$. For a
saturating potential ($\beta > 0$), $\lambda (j)$ is predicted to be {\it
larger} than $\lambda (j=0)$.  For a potential in which the curvature
increases with $x$ ($\beta < 0$) $\lambda (j)$ would be {\it smaller}
than $\lambda (j=0)$. Finally, $\lambda$ should depend upon the {\it
ratio} $j/j_c$ rather than on $j$ alone. We now address each of these
three points experimentally and conclude that in typical high-$T_c$
superconductors, $\lambda (j/j_c)$ always {\em increases} with the
increase of a ratio $j/j_c$ thus providing evidence for a saturating
pinning potential.

The penetration depth was measured with an 11 MHz tunnel-diode driven LC
resonator \cite{resonator,prozorov} operating in a $^{3}$He refrigerator.
An external $DC$ magnetic field (0-7 kOe) was applied using a compensated
superconducting magnet. An $AC$ excitation field ($\sim 5$ mOe) was
applied parallel to the $DC$ field. For platelet superconducting samples
with both $AC$ and $DC$ magnetic fields applied perpendicular to largest
face, an analysis was given in Ref. \cite{prozorov}. When $\lambda$ is
much less than any sample dimension, the resonance frequency shift
$\Delta f = f (T) - f(T_{min})$ of the oscillator is proportional to a
change in the penetration depth, $\Delta \lambda = \lambda (T) - \lambda
(T_{min})$ via $\Delta f = - G \Delta \lambda$, where $G$ is a
calibration constant \cite{resonator,prozorov}. The sample could be moved
in and out of the sense coil, thus allowing for both in situ background
subtraction and calibration \cite{prozorov}. The noise level, $\Delta
f/f_{0}\approx 10^{-9}/\sqrt{Hz}$ permitted a resolution of $\Delta
\lambda _{ac}\leq 0.5$\ \AA\ for typical samples.

The behavior to be described was observed in a variety of superconducting
single crystals including YBaCuO ($T_c \approx 93$ K), BiSrCaCuO ($T_c
\approx 90$ K), PrCeCuO ($T_c \approx 21$ K), and (BEDT) organic ($T_c
\approx$ 12 and 9.5 K) as well as conventional Nb polycrystalline sample
($T_c = 9.27$ K). These properties of the pinning potential persist
despite large differences in $T_c$, symmetry of the order parameter,
anisotropy and the sign of charge carriers.

\begin{figure}
 \centerline{\psfig{figure=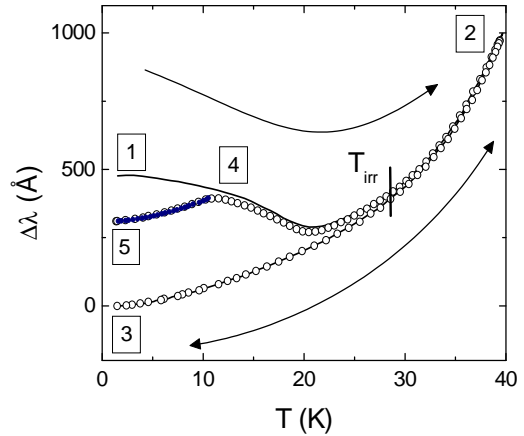,width=7cm,clip}}
 \caption{Curve $1 \rightarrow 2 \rightarrow 3$: $H_{dc}=260$ Oe was
applied at $T=1.5$ K. Then, sample was warmed up and cooled down; Curve
$4 \rightarrow 5$: field was applied at $T=12$ K and sample was cooled
down; Curve $5 \rightarrow 4 \rightarrow 2 \rightarrow 3$: after previous
step, sample was warmed up and finally cooled again to $T=1.5$ K.}
 \label{fig2}
\end{figure}
{\bf Observation of thermal irreversibility:} Figure \ref{fig2} presents
a series of temperature sweeps corresponding to different ratios of
$j/j_c(T)$ for a BSCCO single crystal. In the first experiment (solid
line $1 \rightarrow 2 \rightarrow 3$), the sample was cooled in zero
field to 1.5 K, a $DC$ field of 260 Oe was applied (after first ramping
to -7 kOe to insure complete penetration), and the sample was then warmed
($1 \rightarrow 2$) and then cooled back to 1.5 K ($2 \rightarrow 3$).
Figure \ref{fig1} shows the corresponding trajectory of the system in its
pinning potential well. Once the temperature exceeded a value
$T_{irr}(H)$, $j(t)$ relaxed rapidly and subsequent cooling and warming
traces ($2 \rightarrow 3 \rightarrow 2$) were perfectly reversible. This
curve corresponds to motion near bottom of the pinning well where both
$j$ and $\beta$ are small and $\lambda (T)$ reflects changes in $j_c(T)$
or, equivalently, the height of the well. We refer to the ($2 \rightarrow
3 \rightarrow 2$) curve as the "reversible" curve. It is identical to
that obtained in a direct field-cooled experiment at the same field.
$T_{irr}(H)$ represents a vortex "annealing" or low-$T$ irreversibility
temperature for the described experiment. In our case, $T_{irr}(260 Oe)$
= 28 K, which is close to the temperature at which several changes in
vortex structure are often observed \cite{kes}. This temperature is well
below the usual irreversibility (and melting) temperature determined from
magnetization measurements \cite{kes}.

The irreversible effects are quite small, $\sim 40$ Hz - between marks 1
and 3, Fig. \ref{fig2}. The full frequency shift from the lowest
temperature up to $T_c$ is about $10^{3}$ times larger. The same sample,
measured using a {\it Quantum Design} MPMS SQUID magnetometer, showed a
diamagnetic Meissner signal of $< 8 \times 10^{-5}$ emu. Thus, hysteresis
in $\lambda (T,j)$ is equivalent to a change in magnetization $\leq
10^{-7}$ emu. To resolve the shape of a hysteresis curve a sensitivity
better than $10^{-8}$ emu is needed. The effect is even smaller on
smaller PCCO samples, where a similar estimation gives $\sim 10^{-9}$ emu
for the magnitude of the effect. This sensitivity is difficult to achieve
in commercial AC susceptometers, but measurements of large ceramic
high-$T_c$ samples should, in principle, be able to resolve the
irreversibility.

\begin{figure}
 \centerline{\psfig{figure=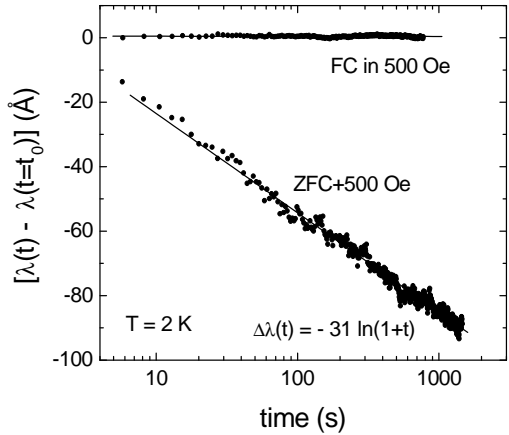,width=7cm,clip}}
 \caption{Time-logarithmic relaxation of the $AC$ penetration depth
after application of a 500 Oe magnetic field at 2 K (lower curve) and
after FC in 500 Oe (upper curve).} \label{fig3}
\end{figure}
{\bf Logarithmic relaxation of $\lambda (j)$:} If Eq. (\ref{lacj}) is
correct, $\lambda (j)$ should relax along with $j$, due to thermally
activated flux creep. Figure \ref{fig3} shows the time dependence of
$\lambda (T,H,t) - \lambda (T,H,t_0)$ at $T=2$ K in BSCCO single crystal.
The sample was cooled in zero field, after which $H_{dc}=500$ Oe was
applied. Since the temperature was constant $j_c(T)$ did not change but
$\lambda (t)$ decreased logarithmically with time. The decrease in
$\lambda (t)$ was independent of the sign of $j$ (i.e., if the field was
reduced after ramping up to a large value). By contrast, the flat curve
shows $\lambda (t)$ after field cooling in 500 Oe to the same
temperature. No noticeable relaxation was observed.

\begin{figure}
 \centerline{\psfig{figure=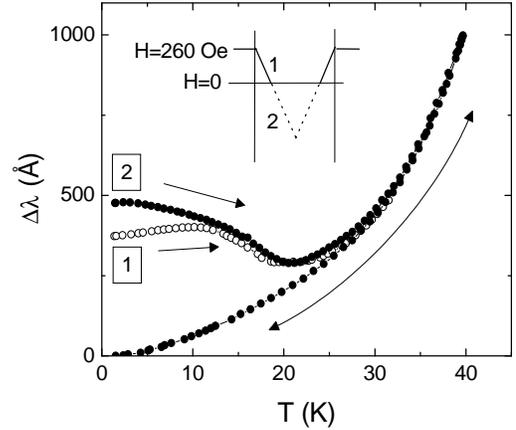,width=7cm,clip}}
 \caption{Curve 1: A 260 Oe magnetic field was applied after ZFC at 1.5
K. Curve 2: After ZFC, magnetic field was first set to -7 kOe an returned
back to 260 Oe. Schematics shows corresponding $DC$ profiles of vortex
density marked $1$ and $2$.}
 \label{fig4}
\end{figure}
{\bf Dependence upon $j/j_c$:} In the next experiment, illustrated in
Fig. \ref{fig2}, the sample was cooled to $T=12$ K and, after ramping to
-7 kOe to ensure a complete flux penetration, a 260 Oe field was applied
(point 4 in Fig. \ref{fig2}) The sample was then {\it cooled} to 1.5 K
($4 \rightarrow 5$) and warmed ($5 \rightarrow 4 \rightarrow 2$) above
$T_{irr}(H)$. Again, upon subsequent cooling and warming ($2 \rightarrow
3 \rightarrow 2$), $\lambda (T)$ remained on the reversible curve and
coincided with the reversible curve obtained in the first experiment
described above. We interpret this measurement as follows. When a $DC$
magnetic field is turned on at $T_{ZFC}=12$ K, $j$ is determined by
$U(j/j_{c}(T)) = T_{ZFC} \ln{(\Delta t/t_0)}$, where $t_0 $ is the
characteristic relaxation time and $\Delta t$ is the experimental
time-window \cite{blatter}. When the temperature decreases, $j_c$
increases and the activation energy $U (j/j_c)$ increases, thus
prohibiting vortices from thermally activated motion and essentially
freezing the value of $j$. If $\lambda$ were dependent only on $j$, it
would have remained constant from mark 4 to mark 5, Fig. \ref{fig2}. The
fact that it decreases implies a parametric change in the pinning
potential: when $j_c$ increases, $j/j_c$ decreases and by Eq. \ref{lacj},
the Campbell length decreases. When the temperature is again increased
($5 \rightarrow 4$) $j(T)$ remains frozen at its value originally set at
$T_{ZFC}$ (mark 4) and $\lambda (T)$ retraces the preceding cooling ($4
\rightarrow 5$) curve reversibly. Above $T_{ZFC}$ K, the activation
energy $U(j/j_{c})$ becomes low enough for $j$ to once again relax. Then,
$\lambda ( T)$ decreases until contribution of the reversible curve
becomes dominant.

{\bf Tilt contribution to the signal, $\lambda_{44}$:} In the
configuration used here (magnetic field normal to the conducting planes
of a thin platelet), the magnetic field penetrates both from sides and
from the top and bottom surfaces. In the Meissner state, this was
demonstrated by solving numerically the London equation and testing the
results experimentally \cite{prozorov}. In the presence of vortices,
penetration from sides, $\lambda_{11}$, is mediated by compressional
waves, whereas penetration from the top and bottom faces is due to tilt
waves, $\lambda_{44}$ \cite{brandt}. For thin enough samples, the latter
contribution must dominate the former as we now show experimentally. In
Fig. \ref{fig4}, the sample was zero-field cooled to $T=1.5$ K and a 260
Oe magnetic field, less than a field of full penetration, was applied
(mark $1$ in Fig. \ref{fig4}). This produced a partially-penetrated state
where only a fraction of the sample interior was filled with vortices, as
illustrated by solid lines in the Bean profile shown in Fig. \ref{fig4}.
Upon warming, $j$ decreased and vortices filled the sample. If most of
the signal comes from the top and bottom surfaces, the penetration depth
must increase as vortices fill the sample. Only when vortices reach the
sample center should the signal decrease in accordance with
Eq.(\ref{lacj}). Exactly this behavior was observed on the curve starting
at point $1$ in Fig. \ref{fig4}. The sample was then cooled from above
$T_c$ in zero field to 1.5 K. H was then ramped down to -7 kOe and then
returned back to +260 Oe, producing the fully penetrated state shown by
the dotted and solid lines in Fig. \ref{fig4}. The resulting curve
beginning at $2$ is shown in Fig. \ref{fig4}. $\lambda (T) $ decreased
monotonically upon warming, indicating that tilt motion of vortices
penetrating the top and bottom surfaces constituted the primary signal
contribution.

\begin{figure}
 \centerline{\psfig{figure=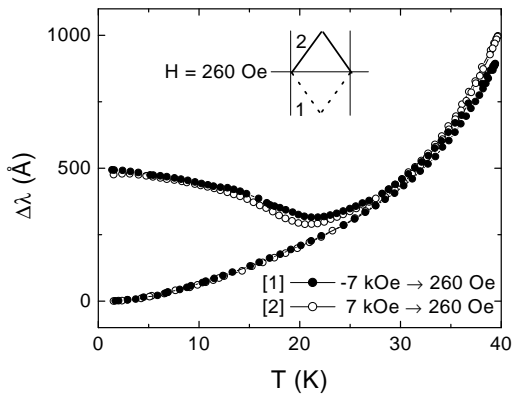,width=7cm,clip}}
 \caption{Comparison of $\Delta \lambda (T)$ for flux entry and exit.
 {\bf Closed symbols:} magnetic field was ramped up from - 7 kOe to +260
Oe (flux entry). {\bf Open symbols:} field was ramped down from +7 kOe to
+260 Oe (flux exit) and the sample was warmed-then-cooled. Schematics
shows the
 corresponding profiles of vortex density.}
 \label{fig5}
\end{figure}
Although the Campbell length in a uniform state depends upon $B$, the
observed hysteresis depends explicitly upon $j/j_c$. (A weak implicit
$B-$dependence may arise through $j/j_c (B)$). This fact is demonstrated
in Fig. \ref{fig5} where we compare $\lambda (T)$ measured at the same
final value of applied field, $H=260$ Oe applied at 1.5 K. The solid
symbols correspond to the initial application of a - 7 kOe magnetizing
field while the open ones correspond to + 7 kOe magnetizing field. Both
fields were then ramped to $H = +260$ Oe before measurements began. These
field changes were sufficient for full flux penetration. The
corresponding flux profiles are shown schematically in Fig. \ref{fig5}.
The value of $B$ throughout the sample was much different for these two
starting conditions, but the thermal hysteresis curves are essentially
identical, demonstrating that the hysteresis in $\lambda_{C}$ is directly
related to $j/j_c$, not B. The small differences at elevated temperatures
could be attributed to the differences in $j/j_c (B)$.

In conclusion, we have demonstrated that thermal hysteresis of the $AC$
penetration depth can be attributed to changes in $j/j_c$ according to
Eq.(\ref{lacj}). The effect is observable only for a non-parabolic
pinning potential that, as follows from our measurements, saturates with
displacement. We identify a new low temperature irreversibility line
above which thermal hysteresis vanishes. A comprehensive study of
$T_{irr}(H)$ in conventional and layered superconductors will be
presented in a forthcoming paper.

Acknowledgments: We thank V. Geshkenbein, A. Koshelev, V. Vinokur, A.
Gurevich, J. R. Clem and E. H. Brandt for useful discussions and
communications. Work at UIUC was supported by Science and Technology
Center for Superconductivity Grant No. NSF-DMR 91-20000. Work at The
University of Tokyo is supported by CREST and Grant-in-Aid for Scientific
Research from the Ministry of Education, Science, Sports and Culture of
Japan. Work at Argonne National Lab supported by U.S. DOE-BES Contract
No. W-31-109- ENG-38 and NSF-STCS Contract No. DMR 91-20000.

\end{multicols}


\vspace{-0.5cm}

\begin{references}
\vspace{-1.5truecm}

\bibitem{campbell} A. M. Campbell and J. E. Evetts, {\it ''Critical
currents in superconductors''} (Taylor \& Francis Ltd., London, 1972).

\bibitem{clem} M. W. Coffey and J. R. Clem, Phys. Rev. Lett. {\bf 67}, 386
(1991); M. W. Coffey and J. R. Clem, Phys. Rev. B {\bf 45}, 10527 (1992).

\bibitem{brandt} E. H. Brandt, Phys. Rev. Lett. {\bf 67}, 2219 (1991);
E. H. Brandt, Physica C {\bf 195}, 1 (1992); E. H. Brandt, Rep. Prog.
Phys. {\bf 58}, 1465 (1995).

\bibitem{koshelev} A. E. Koshelev and V. M. Vinokur, Physica C {\bf 173}, 465 (1991).

\bibitem{beek} C. J. van der Beek, V. B. Geshkenbein, V. M. Vinokur,
Phys. Rev. B {\bf 48}, 3393 (1993).

\bibitem{blatter} G. Blatter, M. V. Feigelman, V. B. Geshkenbein, A. I.
Larkin, and V. M. Vinokur, Rev. Mod. Phys. {\bf 66}, 1125 (1994).

\bibitem{resonator} C. T. Van Degrift Rev. Sci. Inst. {\bf 46}, 599
(1975); A. Carrington {\it et. al.}, Phys. Rev. B {\bf 59}, R14173
(1999).

\bibitem{prozorov} R. Prozorov, R. W. Giannetta, A. Carrington, and F.
M. Araujo-Moreira, cond-mat/0003003.

\bibitem{kes} P. H. Kes {\it et. al.}, J. Phys. I France {\bf 6}, 2327
(1996).

\end{references}
\end{document}